\documentclass{JHEP}
\usepackage{epsfig}
\usepackage{latexsym}
\usepackage{amssymb}
\newcommand{\eref}[1]{(\ref{#1})}

\newcommand{\fref}[1]{Figure~\ref{#1}}
\newcommand{\cref}[1]{Chapter~\ref{#1}}
\newcommand{\bcenter}{\begin{center}}
\newcommand{\ecenter}{\end{center}}
\newcommand{\beq}{\begin{equation}}
\newcommand{\eeq}{\end{equation}}
\newcommand{\bea}{\begin{eqnarray}}
\newcommand{\eea}{\end{eqnarray}}
\newcommand{\ba}{\begin{array}}
\newcommand{\ea}{\end{array}}
\newcommand{\bi}{\begin{itemize}}
\newcommand{\ei}{\end{itemize}}
\newcommand{\bd}{\begin{description}}
\newcommand{\ed}{\end{description}}

\def\IC{\mathbb{C}}

\def\IR{\mathbb{R}}
\def\IF{\mathbb{F}}
\def\IZ{\mathbb{Z}}
\def\IO{\mathbb{O}}

\def\Id{{\rm Id}}

\def\smiley{\hbox{\large$\bigcirc$\hspace{-0.80em}\raise.2ex
\hbox{$\cdot\cdot$}\kern-.61em\lower.2ex\hbox{\scriptsize$\smile$}}\ }
\def\frowny{\hbox{\large$\bigcirc$\hspace{-0.80em}\raise.2ex
\hbox{$\cdot\cdot$}\kern-.635em\lower.2ex\hbox{\scriptsize$\frown$}}\ }

\newcommand{\gen}[1]{\langle #1 \rangle}
\newcommand{\mat}[1]{\left( \matrix{#1} \right)}
\newcommand{\smat}[1]{{\scriptsize \mat{#1}}}
\newcommand{\tmat}[1]{{\tiny \mat{#1}}}
\makeatletter
\let\hangafter\@hangfrom
\makeatother

\makeatletter
\def\overbracket#1{\mathop{\vbox{\ialign{##\crcr\noalign{\kern3\p@}
\downbracketfill\crcr\noalign{\kern3\p@\nointerlineskip}
$\hfil\displaystyle{#1}\hfil$\crcr}}}\limits}
\def\underbracket#1{\mathop{\vtop{\ialign{##\crcr
$\hfil\displaystyle{#1}\hfil$\crcr\noalign{\kern3\p@\nointerlineskip}
\upbracketfill\crcr\noalign{\kern3\p@}}}}\limits}
\def\overparenthesis#1{\mathop{\vbox{\ialign{##\crcr\noalign{\kern3\p@}
\downparenthfill\crcr\noalign{\kern3\p@\nointerlineskip}
$\hfil\displaystyle{#1}\hfil$\crcr}}}\limits}
\def\underparenthesis#1{\mathop{\vtop{\ialign{##\crcr
$\hfil\displaystyle{#1}\hfil$\crcr\noalign{\kern3\p@\nointerlineskip}
\upparenthfill\crcr\noalign{\kern3\p@}}}}\limits}
\def\downparenthfill{$\m@th\braceld\leaders\vrule\hfill\bracerd$}
\def\upparenthfill{$\m@th\bracelu\leaders\vrule\hfill\braceru$}
\def\upbracketfill{$\m@th\makesm@sh{\llap{\vrule\@height3\p@\@width.7\p@}}%
\leaders\vrule\@height.7\p@\hfill
\makesm@sh{\rlap{\vrule\@height3\p@\@width.7\p@}}$}
\def\downbracketfill{$\m@th
\makesm@sh{\llap{\vrule\@height.7\p@\@depth2.3\p@\@width.7\p@}}%
\leaders\vrule\@height.7\p@\hfill
\makesm@sh{\rlap{\vrule\@height.7\p@\@depth2.3\p@\@width.7\p@}}$}
\makeatother

\preprint{UPR-1014-T \\ {\tt hep-th/0210127}}
\title{$G_2$ Quivers}
\author{Yang-Hui He\footnote{This Research was supported in part under
the gracious patronage of the Dept.~of Physics at the University of
Pennsylvania under cooperative research agreement \# DE-FG02-95ER40893
with the U.~S.~Department of Energy.}
\\
Department of Physics,\\
The University of Pennsylvania,\\
209, S.~33rd st.,
Philadelphia, PA 19104-6396\\
\email{yanghe@physics.upenn.edu}
}
\abstract{We present, in explicit matrix representation and a
modernity befitting the community, 
the classification of the finite discrete subgroups
of $G_2$ and compute the McKay quivers arising therefrom. Of physical
interest are the classes of ${\cal N}=1$ gauge theories descending from
M-theory and of mathematical interest are possible steps toward a
systematic study of crepant resolutions to smooth $G_2$
manifolds as well as generalised McKay Correspondences. 
This writing is a companion monograph to 
\href{http://xxx.lanl.gov/abs/hep-th/9811183}{hep-th/9811183} and 
\href{http://xxx.lanl.gov/abs/hep-th/9905212}{hep-th/9905212}, wherein
the analogues for Calabi-Yau three- and four-folds were considered.
}
\keywords{M-theory, Orbifolds, $G_2$ Holonomy, McKay Quivers, D-branes
Probes and Gauge Theory}
\begin{document}
\section{Introduction}
As Ricci-flat threefolds of $SU(3)$ holonomy were upon the centre
stage in the compactification of string theory in the past two
decades, so too have phenomenological considerations lead us in the
past two years to focus on M-theory compactification to real
sevenfolds of $G_2$ holonomy
\cite{Acharya:1999pm,Acharya:2000gb,Atiyah:2000zz,Atiyah:2001qf,0108065,0109152}. And
thus compounded upon our tremendous attention on Calabi-Yau and
K\"ahler manifolds, both physically and mathematically, is our revival
of interest on special holonomy \cite{Vafa,PT} in super string theory
and indeed, on real algebraic geometry.
Inspired thereby, an impressive collection of excellent works appeared
in quick succession within the community. 

Yet, our inherent difficulty with the explicit construction of $G_2$
manifolds has been one fundamental limitation and so far very few
examples of compact $G_2$ sevenfolds are known \cite{Joyce,Joyce2}.

A classic theme in the Calabi-Yau case has been that whereas
understanding string theory on arbitrary members in the space
of these manifolds is substantially challenging, one could gain great
insight by moving to particular points in the moduli space and
notably, to the singular Calabi-Yau's. In other words, we model our
compact (projective) variety with local, non-compact (affine) algebraic
singularities (cf.~\cite{thesis} for some review in this context). 
For example, the Calabi-Yau twofold K3 can be regarded
locally as resolutions of the ALE orbifold $\IC^2/\Gamma$ for $\Gamma$
some discrete finite subgroup - of ADE type - of $SU(2)$, the holomony
of K3.

The advent of the technology of D-brane probes on such singularities
\cite{DM} has brought about systematic methods \cite{LNV} for
constructing low-energy effective gauge theories of various
supersymmetry, matter content and interactions. 
The case of K3 was considered in \cite{JM}, that of local orbifolds
of Calabi-Yau threefolds, in \cite{HanHe,Muto} and that of fourfolds
in \cite{SU4}.

A natural question then arises as to whether we could do the same, as we
did for holonomy $SU(n = 2,3,4)$, for
local $G_2$ and study the discrete finite subgroups $\Gamma$ of
$G_2$ and hence orbifolds of the form $\IR^7/\Gamma$. Yet, {\it
caveat emptur}; for though with the extensive machinery of complex algebraic
geometry, canonical Gorenstein singularities of the form
$\IC^2/(\Gamma \subset SU(2))$, and $\IC^3/(\Gamma \subset SU(3))$ for
abelian $\Gamma$ have been intensively studied and found to admit crepant
resolutions, the same certainly cannot be said for orbifolds of $\IR^7$.
In other words, though in the Calabi-Yau case there are well-studied
resolutions to smooth Ricci-flat manifolds from the said orbifolds
(which in string theory correspond to well-defined
states which become massless at the singular limit), our present lack of
more tools of real algebraic geometry, let alone real
resolutions, hinders a complete understanding of such $G_2$ orbifolds.

Nevertheless, much nice works have been on non-compact $G_2$ manifolds
(cf.~\cite{noncompact} and references therein) and in particular on
D-brane probes on $G_2$ orbifolds
\cite{fst,probe1,probe2,Romel}. Quantum moduli spaces of such theories
have been considered in \cite{Tamar}.
Examples of abelian orbifolds have
been detailed in \cite{fst,Romel} while elegant extensions of $ADE$
singularities to real dimension seven have been addressed in
\cite{probe1,probe2}. Therefore naturally does arise a present want
which desires a immediate supplement and indeed for which \cite{fst} has
kindly beckoned: the classification of the
discrete subgroups of $G_2$.

And thus is the purpose of the present writing. As a companion
monograph to \cite{HanHe,SU4}, we have transcribed some known results,
collected from mathematical works seemingly obscure to the physics
literature \cite{Wales,Cohen,Griess}
and recast them into a compilation explicit in
representation, feasible to computation and abundant in
tabulation. 
These shall constitute Sections 2 and 3. We then calculate
and draw
what we call ``$G_2$ quivers'' from this data in Section 4 and
discuss implications thereof.
%
%=====
%
\section{Some Preparatory Remarks and Nomenclature}
%-------
\subsection{Reducibility and Primitivity}
We are concerned with finite discrete subgroups of Lie groups and
are thus confined to the study of linear transformations, manifesting
as matrix groups acting upon
vector spaces. Indeed, standard in the mathematical
literature is the following terminology which further categorises
such groups. 
The reader is referred to the excellent monographs
\cite{Blichfeldt} and \cite{Yau},  
(or to \cite{SU4} in a context more immediate to this paper), 
for further details.

Essentially, a linear transformation group $\Gamma$ is called 
Intransitive or {\bf Reducible} if it is block-diagonalizable and 
Transitive or {\bf Irreducible} otherwise.
The Irreducible $\Gamma$ can be further divided into 
the {\bf Primitive} and {\bf Imprimitive}, where the imprimitive can
still have blocks of zeros while the primitive groups
generically have no zero entries and are the fundamental building
blocks in the classification.

The usual scheme of classification of the subgroups of Lie groups is
over the field $\IC$ whereas for obvious physical reasons we are
interested in 7-manifolds over $\IR$ and hence discrete subgroups of
$G_2(\IR)$. Henceforth by $G_2$ we shall mean $G_2(\IR)$. Indeed the
classification in light of the categorisations \cite{Wales,Cohen,Griess}
has been performed for
$G_2(\IC)$ and we beg the reader to take heed that in the ensuing
reducibility etc.~refer to the groups over $\IC$. Nevertheless we can
still refer to these groups under their present categories, since
any subgroup of $G_2(\IC)$, being compact, is actually
contained in a maximal compact subgroup of $G_2(\IC)$ and
is hence conjugate to a subgroup $G_2(\IR)$ \cite{Cohen}
and whence the classifications coincide in any event.
Of course,
we shall be careful to take appropriate involutions to ensure that our
matrix groups are indeed real in that they have generators in
$GL(7,\IR)$ and hence have a real 7-dimensional irrep as reflected in
the character tables.

Unless otherwise stated, we adhere to the following nomenclature
throughout the writing. By Lie groups of finite type we mean discrete
finite groups which are the corresponding continuous Lie groups
defined over some Galois field. Thus for example $GL(n;q)$ is the
general linear group $GL(n)$ over the field $\IF_q$; it is thus the
endomorphism for the vector space $\IF_q^{\oplus n}$. By $\Gamma :=
\gen{\{a_i\}}$ we shall mean that the finite group $\Gamma$ is
generated by elements (matrices) $a_i$.
%-------
\subsection{Automorphisms of the Octonions}
Let us first recall some rudimentary facts concerning $G_2$
in light of its linear transformational properties \cite{Wales}.
The Octonions $\IO$ is a real non-associative division algebra.
In particular, it is a 7-dimensional vector space over $\IR$,
endowed with basis $e_0 = \Id$ as well as
$\{e_{i=1,\ldots,7} \}$ satisfying
$$
e_i^2 = -1, \qquad e_ie_j=e_k\mbox{ for } 
        (i,j,k) = (1+r,2+r,3+r) \bmod 7.
$$

On this vector space, a natural quadratic form exists for
any element $x \in \IO = \sum\limits_{i=0}^7 a_i e_i$ (where
$a_i \in \IR$), namely $Q(x) = \sum_i a_i^2$. Thenceforth
the following bilinear and trilinear forms $b(\_,\_)$ and
$t(\_,\_,\_)$ can be established for $x,y,z \in \IO$:
$$
\ba{l}
b(x,y) = \frac12 \left( Q(x+y) - Q(x) - Q(y) \right) \\
t(x,y,z) = b(xy,z)
\ea
$$

In fact, any automorphism of $\IO$ preserves these above forms.
The group Aut$(\IO)$ of these automorphisms is isomorphic to
$G_2(\IR)$, or what we shall refer to\footnote{
        $G_2(\IC)$ is thus Aut$(\IO \otimes_{\IR} \IC)$.}
as $G_2$. 

It is into the discrete finite subgroups of this automorphism
group, as linear transformations of the real vector space
$\IO$, that this writing shall delve.
%
%========
%
\section{The Classification of the Discrete Finite Subgroups of $G_2$}
The classification, in its original form, has been existent in the
mathematical literature for some time \cite{Wales,Cohen}. Such a
result of Wales-Cohen has been transposed into modernity by Griess in
\cite{Griess}. Our first task then, before moving on to quivers and
gauge theories, is to recast yet again, all these marvelous results,
from their perhaps abstruse guise, to a more tangible form,
whose concrete realisation as matrix groups are explicit.

%--------
\subsection{Reducible}
As with all classifications of these discrete finite subgroups of Lie
groups (cf.~\cite{Blichfeldt,Yau}), the reducible groups are always
direct and semi-direct products of Lie subgroups of the parent.
In the case of $G_2$, all these reducible are constructable from the
finite discrete subgroups of $SU(2)$ and $SU(3)$ \cite{Wales,Joyce}.
Explicit representations of these infinite series follow along the
lines of the $ZD$-type groups for $SU(3)$ in \cite{ZD,Muto}, 
composed of two non-commuting pieces, viz.~the $Z$ and the $D$ of $SU(2)
\hookrightarrow SU(3)$, with their generators appropriately
concatenated. 
So likewise could we do so for $G_2$. 

Now in order to preserve the
automorphism structure of $\IO$ and reality of our 7-dimensional
representation, the denouement is that \cite{Wales} only subgroups of
(a) $SU(2) \times SU(2)$ and (b) $SU(3)$ are allowed. Therefore, the
reducible finite subgroups of $G_2$ are quite well-known, as has been
considered in for example \cite{fst,Romel} and easily extended from
the famous ADE subgroups of $SU(2)$ \cite{Klein}
as well as those of $SU(3)$ \cite{Blichfeldt,Yau,HanHe}. Therefore
upon these irreducibles let us not dwell.

%--------
\subsection{Irreducible Imprimitive}
The heart of the classification lies in the irreducibles, which in
some sense reflect the intricacies of the structure of
Aut$(\IO)$. There are in all 7 of these, 2 imprimitive and 5
primitive. To the particulars of these 7 exceptionals let us now
turn. The ensuing computations are done with the extensive aid of
\cite{GAP} to whose writers we are forever indebted.

The first irreducible imprimitive we shall call $II_1$; it has the
following generators
\beq
II_1 :=
\gen{
\tmat{
    0 & 0 & 0 & 0 & 0 & 1 & 0 \cr 0 & 0 & 1 & 0 & 0 & 0 & 0 \cr 0 & 0 & 0 & 
    1 & 0 & 0 & 0 \cr 0 & 1 & 0 & 0 & 0 & 0 & 0 \cr 0 & 0 & 0 & 0 & 1 & 0 & 
    0 \cr -1 & -1 & -1 & -1 & -1 & -1 & 
     -1 \cr 0 & 0 & 0 & 0 & 0 & 0 & 1 \cr  },
\tmat{
    0 & 0 & 0 & 0 & 0 & 1 & 0 \cr 0 & 1 & 0 & 0 & 0 & 0 & 0 \cr 0 & 0 & 1 & 
    0 & 0 & 0 & 0 \cr 1 & 0 & 0 & 0 & 0 & 0 & 0 \cr 0 & 0 & 0 & 0 & 0 & 0 & 
    1 \cr 0 & 0 & 0 & 1 & 0 & 0 & 0 \cr -1 & -1 & -1 & -1 & -1 & -1 & 
     -1 \cr  }
} \ .
\eeq
We see of course that $II_1$ is indeed 7-dimensional over the
reals. This group is in fact none other than the projective
special linear group over the finite field $\IF_7$, $PSL(2;7)$, which
is isomorphic to another Lie group of finite type,
viz.~$GL(3;2)$. Moreover, $II_1$ is a group of order 168 and is in
fact isomorphic to the familiar exceptional group $\Sigma_{168}$ of
$SU(3)$ \cite{Yau,HanHe}.

The character table for $II_1$ is computed as follows.
\[
\ba{cc}
\ba{|c|c|c|c|c|c|c|}
\hline
& 1 & 21 & 24 & 24 & 42 & 56 \cr 
\hline
\Gamma_1 & 1 & 1 & 1 & 1 & 1 & 1 \cr 
\Gamma_2 & 3 & -1 & w & \bar{w} & 1 & 0 \cr 
\Gamma_3 & 3 & -1 & \bar{w} & w & 1 & 0 \cr 
\Gamma_4 & 6 & 2 & -1 & -1 & 0 & 0 \cr 
\Gamma_5 & 7 & -1 & 0 & 0 & -1 & 1 \cr 
\Gamma_6 & 8 & 0 & 1 & 1 & 0 & -1 \cr
\hline
\ea
\qquad
&
\qquad w := \frac{-1-\sqrt{7}}{2}
\ea
\]

Moving on to the next in the irreducible imprimitives, we have $II_2$,
which is generated by
\beq
II_2 :=
\gen{
  \tmat{
    0 & 1 & 0 & 0 & 0 & 0 & 0 \cr 0 & 0 & 1 & 0 & 0 & 0 & 0 \cr 0 & 0 & 0 & 
    1 & 0 & 0 & 0 \cr 0 & 0 & 0 & 0 & 1 & 0 & 0 \cr 0 & 0 & 0 & 0 & 0 & 1 & 
    0 \cr 0 & 0 & 0 & 0 & 0 & 0 & 1 \cr 1 & 0 & 0 & 0 & 0 & 0 & 0 \cr
  } ,
  \tmat{
    0 & 1 & 0 & 0 & 0 & 0 & 0 \cr 0 & 0 & 0 & 1 & 0 & 0 & 0 \cr 0 & 0 & 0 & 
    0 & 0 & 1 & 0 \cr 1 & 0 & 0 & 0 & 0 & 0 & 0 \cr 0 & 0 & 1 & 0 & 0 & 0 & 
    0 \cr 0 & 0 & 0 & 0 & 1 & 0 & 0 \cr 0 & 0 & 0 & 0 & 0 & 0 & 1 \cr
  },
\tmat{ 0 & -1 & 0 & 0 & 0 & 0 & 0 \cr 
     -1 & 0 & 0 & 0 & 0 & 0 & 0 \cr 0 & 0 & 0 & 0 & 0 & 1 & 0 \cr 0 & 0 & 0 &
    -1 & 0 & 0 & 0 \cr 0 & 0 & 0 & 0 & 1 & 0 & 0 \cr 0 & 0 & 1 & 0 & 0 & 0 & 
    0 \cr 0 & 0 & 0 & 0 & 0 & 0 & -1 \cr  } 
} \ .
\eeq

Now $II_2$ is a group of order 1344 and is in fact a central extension of
$II_1$ in the sense that $II_2 / \IZ_2^3 \cong II_1$. It character
table is as follows:
\[
\ba{|c|c|c|c|c|c|c|c|c|c|c|c|}
\hline
& 1 & 7 & 42 & 42 & 84 & 168 & 168 & 192 & 192 & 224 & 224 \cr 
\hline
\Gamma_1 & 1 & 1 & 1 & 1 & 1 & 1 & 1 & 1 & 1 & 1 & 1 \cr 
\Gamma_2 & 3 & 3 & -1 & -1 & -1 & 1 & 1 & w & \bar{w} & 0 & 0 \cr 
\Gamma_3 & 3 & 3 & -1 & -1 & -1 & 1 & 1 & \bar{w} & w & 0 & 0 \cr 
\Gamma_4 & 6 & 6 & 2 & 2 & 2 & 0 & 0 & -1 & -1 & 0 & 0 \cr 
\Gamma_5 & 7 & -1 & -1 & 3 & -1 & -1 & 1 & 0 & 0 &  -1 & 1 \cr 
\Gamma_6 & 7 & 7 & -1 & -1 & -1 & -1 & -1 & 0 & 0 & 1 & 1 \cr 
\Gamma_7 & 7 & -1 & 3 & -1 & -1 & 1 & -1 & 0 & 0 & -1 & 1 \cr 
\Gamma_8 & 8 & 8 & 0 & 0 & 0 & 0 & 0 & 1 & 1 & -1 & -1 \cr 
\Gamma_9 & 14 & -2 & 2 & 2 & -2 & 0 & 0 & 0 & 0 & 1 &  -1 \cr 
\Gamma_{10} & 21 & -3 & -3 & 1 & 1 & 1 & -1 & 0 & 0 & 0 & 0 \cr 
\Gamma_{11} & 21 & -3 & 1 & -3 & 1 & -1 & 1 & 0 & 0 & 0 & 0 \cr
\hline
\ea
\]
with $w$ as in the characters for $II_1$.

Thus concludes the irreducible imprimitives.
%--------
\subsection{Irreducible Primitive}
We now present the irreducible primitives, which, as was mentioned in
the preliminary remarks, are the true fundamental building
blocks. There are 5 of these groups, of substantial size and we
beseech the reader's patience.

The first primitive we shall call $IP_1$, with the rather complicated
generators
\beq
\ba{r}
IP_1 :=
\langle
\frac{1}{c}
  \smat{ 23\,{b1} + 22\,{b2} & 31\,{b1} + 
    41\,{b2} & -15\,{b1} - 12\,{b2} & 7\,
     {b1} - 7\,{b2} & 18\,{b1} + 
    27\,{b2} & 9\,{b2} & 14\,{b1} + 
    22\,{b2} \cr -4\,{b1} - 5\,{b2} & -23\,
     {b1} - 40\,{b2} & 12\,{b1} - 
    3\,{b2} & 7\,{b1} + 20\,{b2} & -9\,
     {b1} - 27\,{b2} & \frac{-c}{2} & -22\,
     {b1} - 32\,{b2} \cr 0 & 0 & 0 & 0 & 0 & 0 & c \cr
    -19\,{b1} - 17\,{b2} & -8\,{b1} - 
    37\,{b2} & -15\,{b1} - 21\,{b2} & 4\,
     {b1} + 23\,{b2} & -45\,{b1} - 
    54\,{b2} & 27\,{b1} + 18\,{b2} & -10\,
     {b1} - 26\,{b2} \cr -5\,{b1} - 
    4\,{b2} & -13\,{b1} - 23\,{b2} & 15\,
     {b1} + 12\,{b2} & -7\,{b1} + 
    7\,{b2} & -9\,{b2} & -9\,{b2} & 4\,
     {b1} - 4\,{b2} \cr 2\,{b1} - 
    2\,{b2} & -2\,{b1} + 2\,{b2} & 12\,
     {b1} + 6\,{b2} & -8\,{b1} - 
    10\,{b2} & 0 & 0 & 2\,{b1} - 
    2\,{b2} \cr 5\,{b1} + 4\,{b2} & 13\,
     {b1} + 23\,{b2} & 3\,{b1} + 
    6\,{b2} & 7\,{b1} - 7\,{b2} & 18\,
     {b1} + 27\,{b2} & -18\,{b1} - 
    9\,{b2} & 14\,{b1} + 22\,{b2} \cr  },
\\
\\
\frac{1}{c}
\tmat{ 5\,{b1} + 4\,{b2} & 13\,{b1} + 
    23\,{b2} & -15\,{b1} - 12\,{b2} & 7\,
     {b1} - 7\,{b2} & 9\,{b2} & 9\,
    {b2} & -4\,{b1} + 4\,{b2} \cr 17\,
     {b1} + 10\,{b2} & {b1} - 
    19\,{b2} & 3\,{b1} + 6\,{b2} & 13\,
     {b1} + 23\,{b2} & -9\,{b2} & -9\,
    {b2} & 8\,{b1} - 8\,{b2} \cr 18\,
     {b1} + 9\,{b2} & 27\,{b1} + 
    54\,{b2} & 9\,{b2} & 9\,{b1} - 
    18\,{b2} & 9\,{b1} + 27\,{b2} & \frac{-c}
    {2} & 18\,{b1} + 36\,{b2} \cr -5\,{b1} - 
    4\,{b2} & -31\,{b1} - 41\,{b2} & 15\,
     {b1} + 12\,{b2} & -7\,{b1} + 
    7\,{b2} & -9\,{b2} & -9\,{b2} & -14\,
     {b1} - 22\,{b2} \cr -19\,{b1} - 
    8\,{b2} & -17\,{b1} - 19\,{b2} & 3\,
     {b1} + 6\,{b2} & -5\,{b1} + 
    5\,{b2} & -36\,{b1} - 27\,{b2} & 18\,
     {b1} + 9\,{b2} & -10\,{b1} - 
    8\,{b2} \cr -19\,{b1} - 17\,{b2} & 10\,
     {b1} + 35\,{b2} & -15\,{b1} - 
    3\,{b2} & -14\,{b1} - 31\,{b2} & -45\,
     {b1} - 18\,{b2} & 27\,{b1} + 
    18\,{b2} & -10\,{b1} + 10\,{b2} \cr 
     {b1} - {b2} & 8\,{b1} + 
    {b2} & -3\,{b1} - 15\,{b2} & -4\,
     {b1} - 5\,{b2} & 9\,{b1} & -9\,
    {b1} & -8\,{b1} - 10\,{b2} \cr  } 
\rangle
\ea
\eeq
with
\[
\ba{l}
x := \cos(\frac{2 \pi}{13}), \qquad c := \frac{9}{1+2x}; \\ 
b1 := 2 - 9x - 2x^2 + 24x^3 - 16x^5, \mbox{~and~} \\
b2 := -3 + 14x + 4x^2 - 44x^3 + 32x^5.
\ea
\]
Now $IP_1$ is nothing but the group $PSL(2;13)$, of order 1092. Its
characters are given below:
\[
\ba{|c|c|c|c|c|c|c|c|c|c|}
\hline
 & 1 & 84 & 84 & 91 & 156 & 156 & 156 & 182 & 182 \cr
\hline
\Gamma_1 & 1 & 1 & 1 & 1 & 1 & 1 & 1 & 1 & 1 \cr 
\Gamma_2 & 7 & p & q & -1 & 0 & 0 & 0 & -1 & 1 \cr 
\Gamma_3 & 7 & q & p & -1 & 0 & 0 & 0 & -1 & 1 \cr 
\Gamma_4 & 12 & -1 & -1 & 0 & r_1 & r_2 & r_3 & 0 & 0 \cr 
\Gamma_5 & 12 & -1 & -1 & 0 & r_2 & r_3 & r_1 & 0 & 0 \cr 
\Gamma_6 & 12 & -1 & -1 & 0 & r_3 & r_1 & r_2 & 0 & 0 \cr 
\Gamma_7 & 13 & 0 & 0 & 1 & -1 & -1 & -1 & 1 & 1 \cr 
\Gamma_8 & 14 & 1 & 1 & -2 & 0 & 0 & 0 & 1 & -1 \cr 
\Gamma_9 & 14 & 1 & 1 & 2 & 0 & 0 & 0 & -1 & -1 \cr
\hline
\ea
\]
with $p := \frac{1-\sqrt{13}}{2}$, $q := \frac{1+\sqrt{13}}{2}$ and
$r_{1,2,3}$ the three roots of the cubic equation
$1 - 2r - r^2 + r^3 = 0$.

The next in the family is $IP_2$, generated by
\beq
IP_2 :=
\gen{
  \tmat{ - \frac{1}{3}   & 0 & 1 & - \frac{4}{3} 
       & \frac{1}{3} & -1 & \frac{5}{3} \cr -3 & -2 & 2 & -1 & 0 & 
     -2 & 3 \cr 0 & 0 & 0 & 0 & -1 & 0 & 0 \cr 0 & 0 & 
     -1 & 0 & 0 & 0 & 0 \cr - \frac{4}{3}   & 0 & 1 & \frac{2}
    {3} & - \frac{5}{3}   & 1 & \frac{2}{3} \cr 4 & 3 & 
     -3 & 2 & 1 & 2 & -4 \cr -1 & 0 & 1 & -1 & 1 & -1 & 2 \cr  },
 \tmat{ - \frac{16}{3}   & -3 & 4 & - \frac{7}{3}
        & - \frac{2}{3}   & -2 & \frac{17}
    {3} \cr 7 & 3 & -6 & 2 & 2 & 2 & -7 \cr - \frac{16}{3}   & 
     -2 & 5 & - \frac{7}{3}   & - \frac{5}{3}   & 
     -1 & \frac{17}{3} \cr 0 & 0 & 0 & 0 & 0 & 0 & 1 \cr - \frac{4}
      {3}   & 0 & 1 & \frac{2}{3} & - \frac{5}{3} 
       & 1 & \frac{2}{3} \cr - \frac{10}{3}   & -2 & 2 & -
      \frac{1}{3}   & - \frac{5}{3}   & -1 & \frac{8}
    {3} \cr 2 & 0 & -3 & 1 & 1 & 0 & -2 \cr  } 
}
\eeq
This group is in fact $PSL(2;8)$, of order 504 and with character
table: 
\[
\ba{|c|c|c|c|c|c|c|c|c|c|}
\hline
    & 1 & 56 & 56 & 56 & 56 & 63 & 72 & 72 & 72 \cr 
\hline
\Gamma_1 & 1 & 1 & 1 & 1 & 1 & 1 & 1 & 1 & 1 \cr 
\Gamma_2 & 7 & -2 & 1 & 1 & 1 & -1 & 0 & 0 & 0 \cr 
\Gamma_3 & 7 & 1 & -p & -q & p + q & -1 & 0 & 0 & 0 \cr 
\Gamma_4 & 7 & 1 & p + q & -p & -q & -1 & 0 & 0 & 0 \cr 
\Gamma_5 & 7 & 1 & -q & p + q & -p & -1 & 0 & 0 & 0 \cr 
\Gamma_6 & 8 & -1 & -1 & -1 & -1 & 0 & 1 & 1 & 1 \cr 
\Gamma_7 & 9 & 0 & 0 & 0 & 0 & 1 & r & s & t \cr 
\Gamma_8 & 9 & 0 & 0 & 0 & 0 & 1 & s & t & r \cr 
\Gamma_9 & 9 & 0 & 0 & 0 & 0 & 1 & t & r & s \cr
\hline
\ea
\]
with $(p,q) := \left( \cos (\frac{4\pi}{9} ), \cos (\frac{8\pi}{9} )
\right)$ and
$(r,s,t) :=  \left( \cos (\frac{2\pi}{7} ), \cos (\frac{4\pi}{7} ),
\cos (\frac{6\pi}{7} ) \right)$.

And thence follows the next imprimitive, $IP_3$, with generators
\beq
IP_3 :=
\gen{
  \tmat{ \frac{20}{7} & 5 & \frac{8}{7} & \frac{10}{7} & - \frac{17}
      {7}   & 6 & - \frac{13}{7} 
       \cr 0 & 0 & 1 & 0 & 0 & 0 & 0 \cr - \frac{1}{7}   & 
     -1 & \frac{8}{7} & - \frac{4}{7}   & \frac{4}{7} & 0 & 
     \frac{1}{7} \cr 0 & 3 & 0 & 1 & -2 & 2 & 
     -1 \cr 0 & 1 & 0 & 0 & 0 & 0 & 0 \cr - \frac{16}{7}   & 
     -4 & - \frac{12}{7}   & - \frac{8}{7}   & 
     \frac{15}{7} & -5 & \frac{9}{7} \cr -4 & -7 & 0 & -1 & 3 & -7 & 1
     \cr  },
  \tmat{ - \frac{33}{7}   & -8 & - \frac{9}{7} 
       & - \frac{13}{7}   & \frac{27}{7} & -9 & \frac{12}
    {7} \cr 0 & 0 & 0 & 0 & 0 & 0 & -1 \cr - \frac{4}{7}   & 
     -1 & - \frac{3}{7}   & - \frac{2}{7}   & 
     \frac{2}{7} & -2 & - \frac{3}{7}   \cr - \frac{5}
      {7}   & 1 & - \frac{2}{7}   & \frac{1}{7} & -
      \frac{1}{7}   & 0 & - \frac{2}{7}   \cr \frac{4}
    {7} & 0 & \frac{3}{7} & \frac{2}{7} & \frac{5}{7} & 1 & - \frac{4}
      {7}   \cr \frac{23}{7} & 5 & \frac{5}{7} & \frac{8}{7} & -
      \frac{15}{7}   & 6 & - \frac{2}{7}   \cr \frac{12}
    {7} & 4 & - \frac{5}{7}   & - \frac{1}{7} 
       & - \frac{13}{7}   & 3 & - \frac{5}{7}
       \cr  }
}
\eeq
This group is isomorphic to $PGL(2;7)$ and is of order 336. Its
characters are:
\[
\ba{|c|c|c|c|c|c|c|c|c|c|}
\hline
& 1 & 21 & 28 & 42 & 42 & 42 & 48 & 56 & 56 \cr 
\hline
\Gamma_1 & 1 & 1 & 1 & 1 & 1 & 1 & 1 & 1 & 1 \cr 
\Gamma_2 & 1 & 1 & -1 & -1 & -1 & 1 & 1 & -1 & 1 \cr 
\Gamma_3 & 6 & -2 & 0 & 0 & 0 & 2 & -1 & 0 & 0 \cr 
\Gamma_4 & 6 & 2 & 0 & -{\sqrt{2}} & {\sqrt{2}} & 0 &  -1 & 0 & 0 \cr 
\Gamma_5 & 6 & 2 & 0 & {\sqrt{2}} & -{\sqrt{2}} & 0 & -1 & 0 & 0 \cr 
\Gamma_6 & 7 & -1 & 1 & -1 & -1 & -1 & 0 & 1 & 1 \cr 
\Gamma_7 & 7 & -1 & -1 & 1 & 1 & -1 & 0 & -1 & 1 \cr 
\Gamma_8 & 8 & 0 & 2 & 0 & 0 & 0 & 1 & -1 & -1 \cr 
\Gamma_9 & 8 & 0 & -2 & 0 & 0 & 0 & 1 & 1 & -1 \cr
\hline
\ea
\]

Our fourth member is the group $IP_4$, generated by
\beq
IP_4 :=
\gen{
  \tmat{ 0 & 0 & 0 & 0 & 0 & 0 & -1 \cr 1 & -1 & 
     -1 & 0 & 0 & 0 & 0 \cr 0 & 0 & 0 & -1 & -1 & 0 & -1 \cr 
     -1 & 1 & 0 & 0 & 0 & 0 & -1 \cr 1 & 0 & 0 & 0 & 0 & 0 & 0 \cr 
     -1 & 0 & 0 & 1 & 1 & -1 & 1 \cr 0 & -1 & 0 & 0 & 1 & 0 & 1 \cr
  },
  \tmat{ -1 & 0 & 0 & 1 & 0 & -1 & 0 \cr 0 & 0 & 0 & -1 & -1 & 0 & -1 \cr 
     -1 & 0 & 0 & 1 & 1 & 0 & 0 \cr 0 & -1 & 0 & 1 & 1 & 
     -1 & 1 \cr 0 & 0 & 0 & 0 & 
     -1 & 0 & 0 \cr 0 & 0 & 1 & 0 & 1 & 0 & 1 \cr 0 & 0 & 0 & 
     -1 & 0 & 0 & 0 \cr  }
}
\eeq

This group $IP_4$ is identified as  $PU(3;3)$, of order 6048. The
character table is:
\[
\ba{|c|c|c|c|c|c|c|c|c|c|c|c|c|c|c|}
\hline
 & 1 & 56 & 63 & 63 & 63 & 378 & 504 & 504 & 504 & 672 & 756 & 756 &
   864 & 864 \cr 
\hline
\Gamma_1 & 1 & 1 & 1 & 1 & 1 & 1 & 1 & 1 & 1 & 1 & 1 & 1 & 1 & 1 \cr 
\Gamma_2 & 6 & -3 & -2 & -2 & -2 & 2 & 1 & 1 & 1 & 0 & 0 & 0 & -1 & 
    -1 \cr 
\Gamma_3 & 7 & -2 & -1 & 3 & 3 & -1 & 0 & 0 & 2 & 1 & -1 & 
    -1 & 0 & 0 \cr 
\Gamma_4 & 7 & -2 & 3 & -1 - 2\,i  & -1 + 
   2\,i  & 1 & -1 - i  & -1 + i  & 0 & 1 & -i  & i
    & 0 & 0 \cr 
\Gamma_5 & 7 & -2 & 3 & -1 + 2\,i  & -1 - 
   2\,i  & 1 & -1 + i  & -1 - i  & 0 & 1 & i  & -i
     & 0 & 0 \cr 
\Gamma_6 & 14 & 5 & -2 & 2 & 2 & 2 & -1 & -1 & 1 & 
    -1 & 0 & 0 & 0 & 0 \cr 
\Gamma_7 & 21 & 3 & 5 & 1 & 1 & 1 & 1 & 1 & 
    -1 & 0 & -1 & -1 & 0 & 0 \cr 
\Gamma_8 & 21 & 3 & 1 & -3 + 2\,i  & -3 - 
   2\,i  & -1 & i  & -i  & 1 & 0 & -i  & i
    & 0 & 0 \cr 
\Gamma_9 & 21 & 3 & 1 & -3 - 2\,i  & -3 + 2\,i  & 
    -1 & -i  & i  & 1 & 0 & i  & -i  & 0 & 0 \cr 
\Gamma_{10} & 27 & 0 & 3 & 3 & 3 & -1 & 0 & 0 & 0 & 0 & 1 & 1 & -1 & 
    -1 \cr 
\Gamma_{11} & 28 & 1 & -4 & -4\,i  & 4\,i  & 0 & i
    & -i  & -1 & 1 & 0 & 0 & 0 & 0 \cr 
\Gamma_{12} & 28 & 1 & -4 & 4\,
   i  & -4\,i  & 0 & -i  & i  & 
    -1 & 1 & 0 & 0 & 0 & 0 \cr 
\Gamma_{13} & 32 & 
    -4 & 0 & 0 & 0 & 0 & 0 & 0 & 0 & -1 & 0 & 0 & \frac{1 - 
     i \,{\sqrt{7}}}{2} & \frac{1 + i \,{\sqrt{7}}}{2} \cr 
\Gamma_{14} & 32 & -4 & 0 & 0 & 0 & 0 & 0 & 0 & 0 & -1 & 0 & 0 & \frac{1 + 
     i \,{\sqrt{7}}}{2} & \frac{1 - i \,{\sqrt{7}}}{2} \cr 
\hline
\ea
\]

Finally, the largest member in our classification, is the group
$IP_5$,
\beq
IP_5 :=
\gen{
  \tmat{ - \frac{5}{2}   & -7 & - \frac{15}{4} 
       & - \frac{1}{4}   & \frac{15}{2} & -2 & \frac{7}
    {2} \cr - \frac{1}{4}   & -3 & - \frac{15}{8} 
       & - \frac{5}{8}   & \frac{13}{4} & - \frac{1}
      {2}   & \frac{3}{4} \cr - \frac{5}{4}   & -5 & 
     - \frac{15}{8}   & \frac{3}{8} & \frac{17}{4} & -
      \frac{3}{2}   & \frac{11}{4} \cr - \frac{3}{4} 
       & 3 & \frac{15}{8} & - \frac{3}{8}   & - \frac{9}
      {4}   & - \frac{1}{2}   & - \frac{3}{4}
        \cr -1 & -5 & - \frac{5}{2}   & - \frac{1}
      {2}   & 5 & -1 & 2 \cr 2 & 2 & \frac{3}{2} & \frac{1}{2} & 
     -2 & 2 & -3 \cr - \frac{1}{4}   & -4 & - \frac{15}
      {8}   & \frac{3}{8} & \frac{17}{4} & - \frac{1}{2} 
       & \frac{3}{4} \cr  } ,
  \tmat{ - \frac{9}{4}   & -4 & - \frac{15}{8} 
       & \frac{3}{8} & \frac{21}{4} & - \frac{3}{2}   & \frac{7}
    {4} \cr - \frac{5}{2}   & -4 & - \frac{9}{4} 
       & \frac{1}{4} & \frac{9}{2} & -2 & \frac{5}{2} \cr \frac{3}
    {2} & 3 & \frac{7}{4} & \frac{1}{4} & - \frac{5}{2} 
       & 1 & - \frac{5}{2}   \cr -1 & -3 & -1 & 0 & 2 & 
     -2 & 3 \cr - \frac{7}{4}   & -3 & - \frac{13}{8}
        & \frac{1}{8} & \frac{15}{4} & - \frac{3}{2} 
       & \frac{5}{4} \cr \frac{1}{2} & 6 & \frac{13}{4} & - \frac{1}
      {4}   & - \frac{11}{2}   & 1 & - \frac{5}
      {2}   \cr - \frac{5}{4}   & 2 & \frac{9}{8} & 
     \frac{3}{8} & - \frac{3}{4}   & - \frac{1}{2} 
       & - \frac{1}{4}   \cr  } 
}
\eeq

This $IP_5$ is in fact none other than $G_2(2)$, i.e., $G_2$ defined
over the Galois field $\IF_2$. The order is the rather formidable
12096 and the character table is
\[
\ba{|c|c|c|c|c|c|c|c|c|c|c|c|c|c|c|c|c|}
\hline
 & 1 & 56 & 63 & 126 & 252 & 252 & 378 & 504 & 672 & 1008 & 1008 & 
   1008 & 1512 & 1512 & 1728 & 2016 \cr 
\hline
\Gamma_{1} & 1 & 1 & 1 & 1 & 1 & 1 & 1 & 1 & 1 & 1 & 1 & 1 & 1 & 1 & 1 & 1 \cr
\Gamma_{2} & 1 & 1 & 1 & 1 & -1 & -1 & 1 & 1 & 1 & -1 & -1 & 1 & 
    -1 & 1 & 1 & -1 \cr 
\Gamma_{3} & 6 & -3 & -2 & -2 & 0 & 0 & 2 & 1 & 0 & 
   -i \,{\sqrt{3}} & i \,{\sqrt{3}} & 1 & 0 & 0 & -1 & 0 \cr 
\Gamma_{4} & 6 & -3 & -2 & -2 & 0 & 0 & 2 & 1 & 0 & i \,{\sqrt{3}} & 
   -i \,{\sqrt{3}} & 1 & 0 & 0 & -1 & 0 \cr 
\Gamma_{5} & 7 & -2 & -1 & 3 & 
    -3 & 1 & -1 & 2 & 1 & 0 & 0 & 0 & -1 & -1 & 0 & 1 \cr 
\Gamma_{6} & 7 & 
    -2 & -1 & 3 & 3 & -1 & -1 & 2 & 1 & 0 & 0 & 0 & 1 & -1 & 0 & 
    -1 \cr 
\Gamma_{7} & 14 & -4 & 6 & -2 & 0 & 0 & 2 & 0 & 2 & 0 & 0 & 
    -2 & 0 & 0 & 0 & 0 \cr 
\Gamma_{8} & 14 & 5 & -2 & 2 & -2 & 2 & 2 & 1 & 
    -1 & 1 & 1 & -1 & 0 & 0 & 0 & -1 \cr 
\Gamma_{9} & 14 & 5 & -2 & 2 & 2 & 
    -2 & 2 & 1 & -1 & -1 & -1 & -1 & 0 & 0 & 0 & 1 \cr 
\Gamma_{10} & 21 & 3 & 5 & 1 & -1 & 3 & 1 & -1 & 0 & -1 & -1 & 1 & 1 & 
    -1 & 0 & 0 \cr 
\Gamma_{11} & 21 & 3 & 5 & 1 & 1 & -3 & 1 & 
    -1 & 0 & 1 & 1 & 1 & -1 & -1 & 0 & 0 \cr 
\Gamma_{12} & 27 & 0 & 3 & 3 & 
    -3 & -3 & -1 & 0 & 0 & 0 & 0 & 0 & 1 & 1 & -1 & 0 \cr 
\Gamma_{13} & 27 & 0 & 3 & 3 & 3 & 3 & -1 & 0 & 0 & 0 & 0 & 0 & -1 & 1 & 
    -1 & 0 \cr 
\Gamma_{14} & 42 & 6 & 2 & -6 & 0 & 0 & 
    -2 & 2 & 0 & 0 & 0 & 0 & 0 & 0 & 0 & 0 \cr 
\Gamma_{15} & 56 & 2 & 
    -8 & 0 & 0 & 0 & 0 & -2 & 2 & 0 & 0 & 0 & 0 & 0 & 0 & 0 \cr 
\Gamma_{16} & 64 & -8 & 0 & 0 & 0 & 0 & 0 & 0 & 
    -2 & 0 & 0 & 0 & 0 & 0 & 1 & 0 \cr
\hline
\ea
\]

And so these above are the groups of our interest, the 7 exceptional
discrete finite irreducible subgroups of $G_2$.
Thus armed, let us now venture into the physics.
%%%%%%%%%%%%%%%%%%%%%
%================
%%%%%%%%%%%%%%%%%%%%%
\section{McKay Quivers, $G_2$ and Gauge Theories}
The construction of world-volume gauge theories for D-brane probes
transverse to orbifold singularities was pioneered in \cite{DM}
wherein the plethora of mathematical machinery in the resolution of
singularities using the hyper-K\"{a}hler quotient
was brought into string theory. The case then studied, as well as by
subsequent works, have been on local models of non-compact Calabi-Yau
manifolds. In other words the orbifolds originated from the discrete
finite subgroups of the holonomy $SU(2)$ (cf.~\cite{DM,JM}), $SU(3)$
(cf.~\cite{HanHe,Yau,Muto}) and $SU(4)$ (cf.~\cite{SU4,Blichfeldt}).

The general methodology for retrieving the world-volume gauge data
from orbifolds was outlined in \cite{LNV} where the D-terms and
F-terms of the gauge theory became purely dependent upon the
group-theoretical properties of the orbifold, notably the group
representation ring and in particular the Clebsch-Gordan composition
therein. Recently, \cite{fst} and \cite{Romel} discussed some abelian
examples of $G_2$ orbifolds in this context.

We here study M-theory
in the back-ground of a singular spacetime: to
result in ${\cal N}=1$ supersymmetry in four dimensions it is
well-known that one needs to ``compactify'' on real sevenfolds of
$G_2$ holonomy. Whence our singularity $X$ will be non-compact local $G_2$
of the form $\IR^7 / (\Gamma \subset G_2)$ and we are studying some
brane probe on $X$. Of course the resolution of
singularities of this type, especially crepant ones to $G_2$ manifolds
still remain an open question in mathematics \cite{Joyce}
due largely to the want of more powerful techniques in {\em real}
algebraic geometry. Yet let us trudge on.
\subsection{D-Brane Probes in Type II}
Following the prescription of \cite{fst,probe1,probe2}, we will study
the D-brane probe theories of type II originating from the parent
M-theory. Indeed, we have much more knowledge of D-brane worldvolume
technology than M-branes. Therefore our natural setting will be the
reduction of the M-brane probe theory on the singular space $X$ of
$G_2$ holonomy, to a D-brane probe theory of type IIA.

As pointed out in \cite{probe1}, we can do so in two ways. We can
reduce on an $S^1$ transverse to both the M-probe and to
$X$. In this case we have $X$ being the
Higgs branch of the moduli space of the D2-brane world-volume gauge
theory. Alternatively, we can reduce on an $S^1$ contained within $X$,
leading to type IIA backgrounds with D6-branes and/or RR flux which
may give rise to extra subtleties. In this case
$X$ will be the Coulomb branch of the D2-brane
theory. We will focus on the first construction of the Higgs branch.

Therefore the preparatory work in the above sections will be in
service to the ${\cal N}=1$ SUSY theory in three dimensions on the
D2-probe. Mirror to this picture is the type IIB perspective of
\cite{fst} wherein one has a D1-probe and the world-volume theory is
$(1,1)$ sigma model in two dimensions.

The extraction of the matter content and interactions follow the
canonical methods mentioned above \cite{DM,LNV,HanHe}
and we shall see the natural
emergence of the McKay quiver \cite{McKay}.
\subsection{World-Volume Theories and McKay Quivers}
The matter content descents from the parent theory of the 
D-brane in flat space and the resulting bi-fundamentals are
summarised by a quiver diagram whose adjacency matrix $a_{ij}$ is
determined as
\beq
\label{matter}
R^{(7)} \otimes R^{(i)} = \bigoplus\limits_j a_{ij} R^{(j)} \ ,
\eeq
where $R^{(i)}$ is the $i$-th irreducible representation of the
orbifold group $\Gamma$ and $R^{(7)}$ is the defining 7-dimensional
representation which for us is a real $7 \times 7$ matrix. Indeed we
choose the 7-dimensional irrep so as to guarantee that our orbifold
action resides in the full $G_2$ and not any subgroup thereof, such as
$SU(3)$ (which would make our space essentially a Calabi-Yau threefold).
We refer
the reader to \cite{LNV} for the details, 
\cite{HanHe} for a summary and \cite{fst} for the present guise of
the derivation of \eref{matter}.

To \eref{matter} shall the character tables of the previous section
lend an immediate hand: we can instantly invert the equation to arrive
at the quiver \cite{HanHe} as
\beq
\label{quiver}
a_{ij} = \frac{1}{g}\sum\limits_{\gamma
=1}^{r}r_{\gamma }\chi_{\gamma }\chi_{\gamma }^{(i)}\chi
_{\gamma }^{(j)*}
\eeq
where $\chi_{\gamma }^{(i)}$ is the $i$-th irreducible
character for the conjugacy class
represented by $\gamma \in \Gamma$ and $\chi_{\gamma }$ is the
character of our chosen defining 7-dimensional real
irrep. Furthermore, $g=\left| \Gamma \right| $ is the
order of the orbifold group $\Gamma$ and  $r_{\gamma }$ is the order
of the conjugacy class of $\gamma$. The sum extends
over the $r$ conjugacy classes, which by the orthogonality theorem of
characters is equal to the number of irreps.

Therefore standard results dictate that
if we have $n$ parallel coincident branes (in the regular
representation $n=Ng$), then from the
parent $U(n)$ SYM would result a daughter gauge theory which has gauge
group $\prod\limits_i^r U(N n_i)$ with $a_{ij}$
bifundamentals transforming in the $U(N n_i) \times U(N n_j)$ factor.

We see of course, that $a_{ji} = \frac{1}{g}\sum
r_{\gamma }\chi_{\gamma }\chi_{\gamma }^{(j)}\chi_{\gamma }^{(i)*}$
which since the adjacency matrix has integer entries must equal to
$a_{ji}^* = \frac{1}{g}\sum
r_{\gamma }\chi_{\gamma}^*\chi_{\gamma }^{(j)*}\chi_{\gamma
}^{(i)}$. The latter is equal to $a_{ij}$ precisely because our
defining representation is real and thus $\chi_{\gamma}^* =
\chi_{\gamma}$. Hence $a_{ji} = a_{ij}$ and our quivers are
symmetric. In other words the reality of our singularity in the sense
that the orbifold is a real algebraic variety compels us not to have
chiral matter and we have a ``non-chiral'' ${\cal N}=1$ theory in three
dimensions\footnote{We are of course being cavalier with the word
	chiral which for phenomenological purposes are of interest to
	four dimensional theories; by non-chiral here we merely mean
	non-oriented quiver and ergo symmetric $a_{ij}$.}. 
In order to arrive at chiral fields, one must use the
complex 7-dimensional representation for $R^{(7)}$, yet this is
geometrically less clear and complications shall arise as to how one
finds a real $G_2$ locus in the complex 7-dimensional quotient.

\EPSFIGURE[ht]{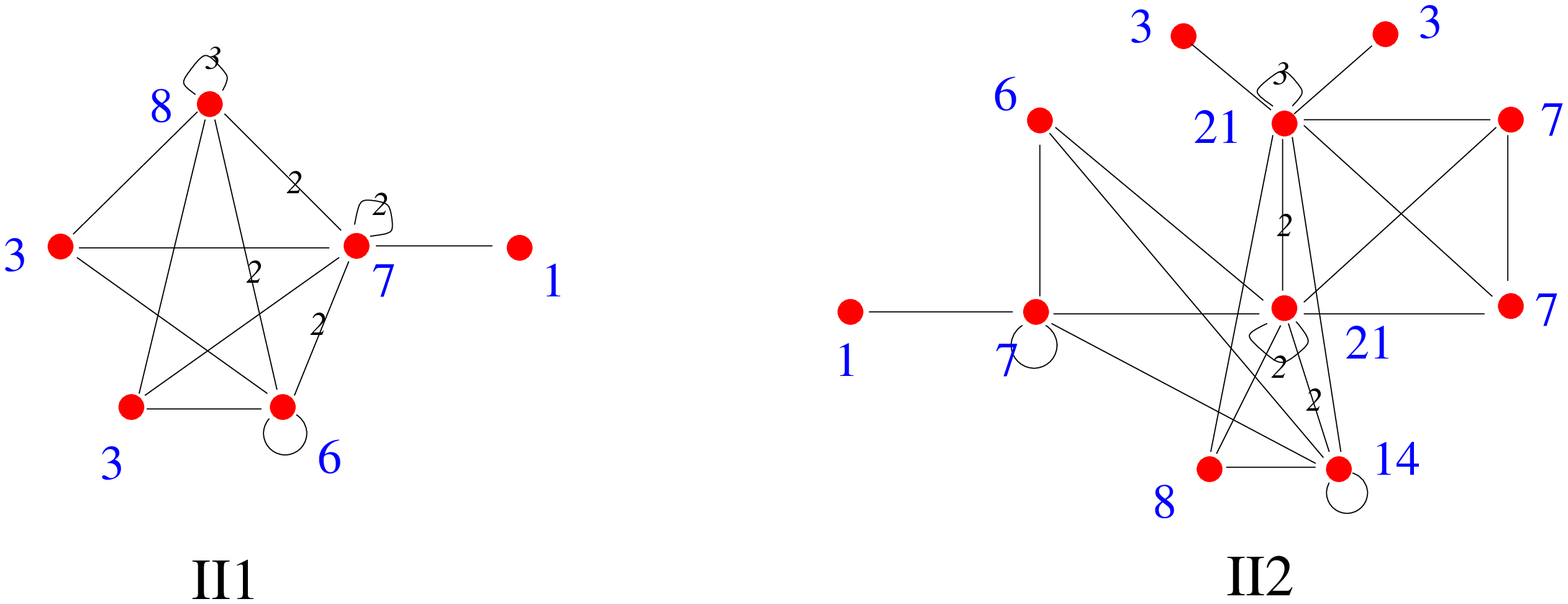,width=5.5in}
{The quiver diagram with respect to the fundamental 7 for the
2 irreducible imprimitive discrete subgroups of $G_2$.
\label{f:II}
}

The McKay quivers of
\eref{quiver} nevertheless provides us with an interesting class of
non-oriented finite graphs. We present them in \fref{f:II} for the
imprimitives and \fref{f:IP} for the primitives. The labels of the
nodes (in blue) are the $n_i$'s in the abovementioned product gauge
group $\prod\limits_i^r U(N n_i)$. In the case of $\Gamma \subset
SU(2)$ these
labels, by virtue of the McKay Correspondence \cite{McKay}, 
are precisely the dual
coxeter numbers of of the affine ADE Dynkin diagrams. Moreover, each
edge in the graph is a bi-directional arrow due to the symmetry
(non-chirality) of the adjacency matrix; multiplicities of these
arrows are indicated thereupon.

\EPSFIGURE[ht]{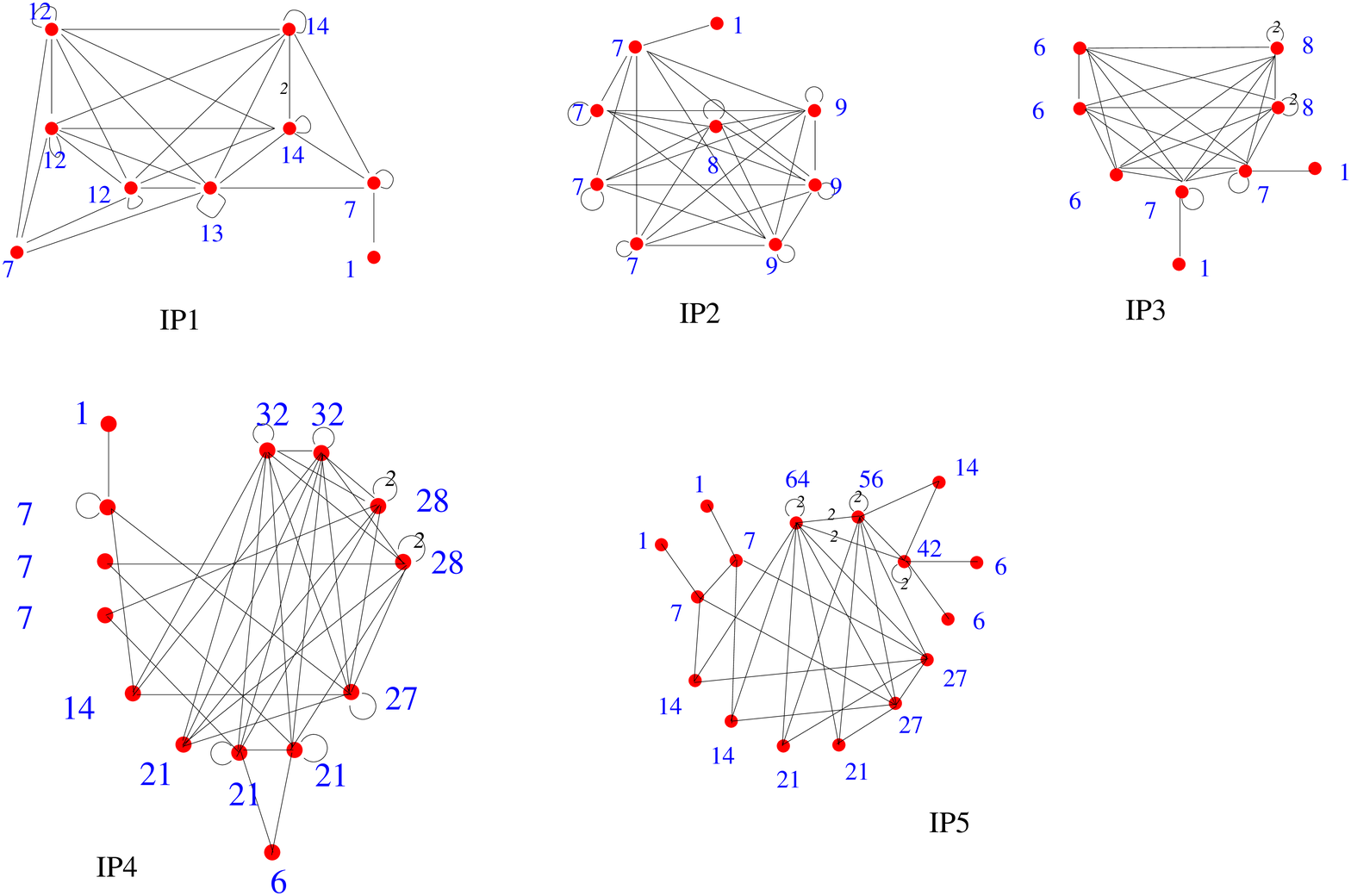,width=7.2in}
{The quiver diagram with respect to the fundamental 7 for the
5 irreducible primitive discrete subgroups of $G_2$.
\label{f:IP}
}

As a parting digression let us briefly comment on some implications of
these graphs. Indeed, as brane-probe theories, each graph corresponds
to an ${\cal N}=1$ gauge theory
whose superpotential could also be
computed using the Clebsch-Gordan coefficients of the respective
groups in the manner of \cite{LNV}. 

Furthermore, it was conjectured in
\cite{HanHe} and addressed further in \cite{HeSong,Finite,0009077}
(see \cite{thesis} for some review) that string orbifolds provides
some type of generalised McKay Correspondence between the
representation ring of discrete
subgroups of $SU(n)$ and the fusion ring of $\widehat{su(n)}$
Wess-Zumino-Witten models at least for $n=2,3,4$ where D-brane probe
technology is applicable. So these were the cases for Calabi-Yau
orbifolds, now we have $G_2$-orbifolds and M-theory; could there be
similar relations to $\widehat{g_2}$ WZW models? Indeed, some
exceptional cases for the latter model were found at levels 3 and 4
\cite{modG2} while the complete classification is still in want; could
these perhaps be in correspondence with the quivers thus far presented?
%%%%%%%%%%%%%%%%%%%%%
%================
%%%%%%%%%%%%%%%%%%%%%
\section*{Acknowledgements}
{\it Ad Catharinae Sanctae
Alexandriae et Ad Majorem Dei Gloriam...\\}
I gratefully acknowledge T.~Gannon of the University of Alberta for and
F.~L\"ubeck of the Lehrstuhl D f\"ur Mathematik, 
RWTH Aachen for enlightening correspondences and informing me of the
various particulars of the literature. 
I would also like to extend my
gratitude to S.~Kachru who many years ago engaged my interest in this
problem as well as A.~Hanany and N.~Prezas for reviving
that interest. Indeed I am much indebted to B.~Feng and
A.~Hanany for valuable comments and review of the draft, especially to
the latter who still like a kind mentor takes me under his 
wings.

We thank D.~Tong and J.-B.~Zuber for useful comments
on the early versions of the preprint.
%%%
%%%
%%%
\bibliographystyle{JHEP}

\end{document}